\documentstyle[12pt,psfig,epsf]{article}

\def\be{\begin{equation}}
\def\bea{\begin{eqnarray}}
\def\ee{\end{equation}}
\def\eea{\end{eqnarray}}
\def\ri{\rightarrow}

\def\RN{{R\over N}}
\def\ra{\rangle}
\def\la{\langle}
\def\r{\right}
\def\l{\left}

\def\rh{\rho}

\def\a{\alpha}
\def\t{\tau}
\def\g{\gamma}
\def\s{\sigma}
\def\b{\beta}

\begin{document}
\title{Relevant Elements, Magnetization and Dynamical Properties in Kauffman
Networks: a Numerical Study.}
\author{U. Bastolla$^{1,}$ $^2$ and G. Parisi$^1$}
\maketitle
\centerline{$^1$Dipartimento di Fisica, Universit\`a ``La Sapienza'', P.le Aldo
Moro 2, I-00185 Roma Italy}
\centerline{$^2$HLRZ, Forschungszentrum J\"ulich, D-52425 J\"ulich Germany}
\medskip
\date{}
\medskip
\centerline{Keywords: Disordered Systems, Genetic Regulatory Networks,}
\centerline{Random Boolean Networks, Cellular Automata}

\begin{abstract}
This is the first of two papers about the structure of Kauffman networks.  
In this paper we define the relevant elements of random networks of 
automata, following previous work by Flyvbjerg \cite{F} and Flyvbjerg and 
Kjaer \cite{FK}, and we study numerically their probability
distribution in the chaotic phase and on the critical line of the
model. A simple approximate argument predicts that their number
scales as $\sqrt N$ on the critical line, while it is linear with $N$ in
the chaotic phase and independent on system size in the frozen phase.
This argument is confirmed by numerical results. The study of the relevant
elements gives useful information about the properties of the
attractors in critical networks, where the pictures coming from either
approximate computation methods or from simulations are not very clear.
\end{abstract}

\section{Introduction}
Kauffman networks are disordered dynamical systems proposed by Kauffman in
1969 as a model for genetic regulatory systems \cite{K}. They
attracted the interest of physicists in the  80's \cite{DP,DW,DF1,F,FK},
due to their analogy with the disordered systems studied in
statistical mechanics, such as the mean field Spin Glass \cite{MPV}. A
dynamical phase transition was found and studied in the framework of
mean field theory.

In this and in the next paper \cite{BP3} we deal with some structural
properties of the networks that determine their attractors. In the
present paper we introduce the relevant elements, a notion that was
suggested by Flyvbjerg \cite{F} and Flyvbjerg and Kjaer \cite{FK}, and
we study their probability distribution. In the next one we describe
how the relevant elements are subdivided into asymptotically non
communicating, independent modules. The modular organization of random
boolean networks was already suggested by Kauffman \cite{K}, and it was
used by Flyvbjerg and Kjaer to study analytically the attractors
in $K=1$ networks. We shall show that it is possible to describe the
phase transition in random boolean networks in terms of the scaling of
the number of relevant elements with system size, or in terms of a
percolation transition in the set of the relevant elements. The
interest of this approach is that some consequences about the
statistical properties of the attractors can be directly drawn. 

In \cite{BP0} we computed the properties of the attractors in the
framework of the annealed approximation, introduced by Derrida and
Pomeau \cite{DP}, but we observed that the results of this
approximation are reliable only when the system is chaotic enough,
becoming exact for a random map. The study of the relevant elements is
complementary to this approach, and we sketch the lines of a new
approximation scheme that works better in the frozen phase and on the
critical line. This region in parameter space is the most interesting
one, since, according to Kauffman, it reproduces some features of real
cells, and is also the less understood, since neither approximate
computations nor simulations \cite{BP1} give a precise picture of the
properties of the attractors for systems of large size.

In next section we define the model, discussing some old results
together with open problems.  In section 3 we define the relevant 
elements and in section 4 we give an approximate argument predicting
the scaling of their number with system size in the different phases
of the model. In the following section we present our numerical
results, starting from the magnetization and the stable elements
(section 5.1) and then discussing the distribution of the relevant
elements and its connection with the properties of the attractors,
respectively in the chaotic phase (section 5.2) and on the critical line
(section 5.3). The discussion of the results is postponed to our
following paper \cite{BP3}, concerning the modular organization of the
relevant elements on the critical line.

\section{Definition of the model and previous works}

Kauffman model is defined as follows. We consider a set of $N$ elements
$\Omega=\{1,\cdots N\}$ and we associate to each of them a  binary
variable, $\s_i\in\{0,1\}, i\in\Omega$. In the biological interpretation
proposed by Kauffman each element of the network represents one gene
and the binary variable $\s_i$ represents its state of activation.

Each element is under the control of $K$ elements, in the sense that
its state at time $t+1$ is determined by the states at time $t$ of the
$K$ control genes, $j_1(i),\cdots j_K(i)$ and by a response function
of $K$ binary variables, $f_i(\s_1,\cdots \s_K)\in \{0,1\}$, that
specifies how the element $i$ responds to the signals coming from
its control variables. The control elements are chosen in $\Omega$
with uniform probability. The response functions are also extracted at
random, and it's believed that the properties of the model do not
depend on the details of their distribution
\cite{F,BP0,BP1}. The rule most generally used in teh literature is
the following: for each of the possible inputs $\in \{0,1\}^K$ we
extract independently the value of $f_i$, and we call $p$ the
probability that $f_i$ is equal to 0.

The dynamics of the system obey the equation

\be \s_i(t+1)=f_i\l(\s_{j_1(i)},\cdots \s_{j_K(i)}\r). \ee

This evolution law is deterministic, but the system is disordered
because the control rules (elements and functions) are chosen at
random from the beginning and kept fixed: thus we deal with a
statistical ensemble of deterministic dynamical systems, and we are
interested in the statistical properties of systems of large size.
For finite $N$, every trajectory becomes periodic after a
long enough transient time, and the configuration space is partitioned into the
attraction basins of the different periodic orbits. We are interested
in the probability distributions of the number, the length and the size
of the attraction basin of the periodic orbits, as well as in that of
transient times. In the biological metaphor, given a set of rules (a
genome) an attractor represents a possible cellular type, its length
represents the duration of the cellular cycle, and the number of
attractors represents the number of cells that can be formed with a
given genome. 

It was observed already in the first simulations that two dynamical regimes
are present, and that the line separating them has properties
reminiscent of those of real cells \cite{K}. In the so-called chaotic phase
(large connectivity, $p$ close to $1/2$) the average length of the
cycles increases exponentially with system size. The limit case of the
chaotic phase, $K\ri\infty$, was already known as Random Map in the
mathematical literature, and was studied in detail by Derrida and
Flyvbjerg \cite{DF2}, who pointed out interesting analogies between this
system and the mean field Spin Glass \cite{MPV} concerning the
distribution of the weights of the attraction basins. In the frozen phase,
on the other hand, the typical length of the cycles does not increase with
$N$. The limit case of this phase, $K=1$, was analytically studied to
some extent by Flyvbjerg and Kjaer \cite{FK}, who introduced in that
context the concept of relevant elements (though without using this name).

The first description of this dynamical phase transition in terms of
an order parameter was given by Derrida and Pomeau \cite{DP}. They
studied the evolution of the Hamming distance between configurations in
the Kauffman networks approximating it with a Markovian stochastic
process. Such approximation (the so-called annealed approximation) was
then shown to be exact in the infinite size limit, concerning the
average value of the distance \cite{DW}. Below a critical line in
parameter space the average distance goes to zero in the infinite size
limit ({\it frozen phase}) and above it the distance goes
to a finite value ({\it chaotic phase}). The position of the
phase transition depends only on the parameter $\rh$,
representing the probability that the responses to two different signals are
different\footnote
{In terms of $p$ one has $\rh=2p(1-p)$, so its value is comprised
between zero and 1/2, but for $K=1$  $\rh$ can be taken as an
independent parameter in [0,1]}, and is given by the equation $\rh_c(K)=1/K$.

The properties of the attractors can be easily computed from the
knowledge of the whole stationary distribution of the distance, and
this can also be obtained within the annealed approximation \cite{BP0}, but the
validity of this approximation in this more general case is not
guaranteed. Comparison with simulations shows that the agreement is
satisfactory in the chaotic phase, while the approximation fails on
the critical line. In the chaotic phase it is possible to compute
the value of the exponent of the typical length of a cycle,
$\t\propto\exp\l(\a(K,\rh) N\r)$, in good agreement with numerical
results, but the distribution of cycle lengths is much
broader than it is expected. The annealed approximation predicts also
that the distribution of the weights of the attraction basins
is universal in the whole chaotic phase, and equal to the
one obtained by Derrida and Flyvbjerg in the case of the Random Map
\cite{DF2}. The corrections to this prediction appear small, if
any, even for $K=3$. Finally, the number of different
cycles in a network is expected to be linear in  $N$, but it is very
hard to test numerically this prediction.

The annealed approximation makes also predictions about the critical
line of the model \cite{BP0}. It predicts that the properties of the
attractors are universal on the critical line $\rh=1/K$ (with the
exceptions of the points $K=1, \rh=1$ and $K=\infty, \rh=0$, which are
not transition points). In particular, the typical length of the cycles
should increase as $\sqrt N$ all along the critical line. Numerical
results are not clear under this respect \cite{BP1}: it seems that the
rescaled cycle length $l=L/\sqrt N$ has a limit distribution if $l$ is
small (roughly, smaller than 2) but for larger values the distribution
becomes broader and broader as $N$ increases \cite{Bhatta,BP1}, so
that it is possible to define an effective length scale increasing
much faster with system size (as a stretched exponential). The
distribution of the number of cycles has exactly the same
characteristics. These results cast doubts on the validity of the
biological analogy proposed by Kauffman, that relies very much
on the fact that in critical networks the typical number of cycles
scales as $\sqrt N$, reminiscent of the fact that the number of cell
types of multicellular organisms very far apart in the filogenetic
tree scales as the square root of the number of the genes, and that in
critical networks the typical length of the cycles increases as a
power law of system size, also consistently with the behavior of cell
cycles time. Thus it is interesting to understand how these
distributions look like in the limit of very large systems.

Another reason of interest of the present approach is that it allows
to understand the limits of the annealed approximation. In our
interpretation the annealed approximation is valid as far as the system loses
memory of the details of its evolution. This, of course, does not
happen if in a realization of a random network some structural
properties that are able to influence its asymptotic dynamics
emerge. Thus the approach presented here is complementary to the one
used in \cite{BP0}.

\section{Definition of the relevant elements}




Let us start recalling the definition of the stable elements \cite{F}. 
These are elements that evolve to a constant state,
independent of the initial configuration. Flyvbjerg defined them and
computed their fraction $s=S/N$ using the annealed approximation, which
becomes exact in the infinite size limit. We now recall briefly, for
future convenience, the main steps of this calculation.

Let us suppose that an element is controlled by $K-i$ stable elements
and $i$ stable ones. Then it will be stable if the control function
does not depend on the unstable arguments when the stable arguments
assume their fixed values. Otherwise it will be unstable. When all the
$i$ unstable elements are different (this can always be taken to be the
case if $K$ is finite and $N$ grows), the probability $P_i$ to
choose a constant control function of $i$ binary variables is given by
$P_i=p^{n_i}+(1-p)^{n_i}$, with $n_i=2^i$. In the framework of the
annealed approximation, extracting at random connections and response functions
at each time step, we get the following equation for the fraction of variables
that are stable at time $t$:
\be\label{stable}
 s(t+1)=\gamma\l(s(t)\r)=
\sum_{i=0}^K {K\choose i} s(t)^{K-i}\l(1-s(t)\r)^i P_i. \ee

This equation can be shown to be exact in the infinite size limit.
The fixed point of this map (which can be interpreted as a
self-consistency equation for the fraction of stable variables) has only the
trivial solution $s=1$ in the frozen phase, in other words all the
elements are stable except eventually a number increasing less than
linearly with $N$. In the chaotic phase this solution becomes unstable
and another solution less than 1 appears. This happens when
$K(1-P_1)=1$. Since $1-P_1=\rh$ (it is just the probability that the
response to two different signals are different) this condition is
equivalent to the condition obtained from the study of the Hamming distance.

The existence of the stable variables is due to the finite
connectivity of the network ($s^*$ goes to zero very fast when $K$
increases). These variables do not take part in the asymptotic
dynamics. Among the remaining unstable variables, some are
irrelevant for the dynamics, either because they do not send signals to
any other variable, or because they send signals, but the response
functions are independent of this signal when the stable variables
have attained their fixed values. The remaining variables, that are
unstable and control some unstable variable, are what we
call the relevant variables. They are the only ones that can influence
the long time behavior of the system.

To be more clear we now describe the algorithm that we used to
identify the relevant variables.
As a first step, we have to identify the stable
variables. These are the variables that assume the same constant state
in every limit cycle, and identifying them is computationally very
hard, but very simple in principle. We then eliminate from the system
the stable variables, reducing the response functions to functions of
the unstable variables alone. Some of the connections left are still
irrelevant, and we have to eliminate them (a connection between the
elements $i$ and $j$ is irrelevant if the reduced response function
$f_i(\s_{j_1(i)},\cdots \s_{j_{K_i}(i)})$ does not depend on the argument
$\s_j$ for all the configurations of the remaining $K_i-1$ control variables).
At this point we iterate a procedure to eliminate the irrelevant
variables. At each iteration we eliminate the variables that do not
send any signal to anyone of the variables that are left, until we
remain with a set that cannot be further reduced. This is the set of
the relevant variables.

Measuring the number of relevant variables is computationally a
very hard task. In order to identify the stable variables, in fact, we
should find all the cycles in the network, and, to be rigorous, we
should simulate a number of trajectories of the same order of the
number of configurations in the system. Of course this is not
feasible and we run only 200 (in some case 300) randomly chosen
trajectories in every network. Thus we overestimate the
number of stable elements. Nevertheless, the number of stable elements
changes very little when we simulate more initial conditions and we think
that the error that we make is not very large. However, for every
network we simulate some hundreds of trajectories and every
trajectories has to be followed until the closing time. This grows
exponentially with system size in the chaotic phase. On the critical
line the typical closing time increase roughly as a power law of
system size, but the distribution becomes broader and broader and the
average closing time is more and more dominated by rare events. The
average depends thus on the number of samples generated and on the
cutoff of the closing time, {\it i.e.} the maximum time that we are
disposed to wait to look for a cycle. To reduce the bias determined by
the cutoff, we had to run simulations lasting a time which increases
roughly as a stretched exponential of system size on the critical
line. Thus it is not possible to simulate systems of more than about
one hundred elements in the chaotic phase and one thousands of
elements on the critical line.

\section{Scaling argument in the frozen phase}

The mean field analysis \cite{F} shows that the fraction of
relevant variables vanishes in the frozen phase and on the critical line,
but does not tell how the number of relevant variables scales with $N$ as $N$
grows. In order to clarify this point, we have to go beyond the mean
field picture.

In the special case of $K=1$, belonging to the frozen phase for every
$\rh <1$, there are detailed analytical results about the distribution
of the relevant variables \cite{FK}. We propose here a rough argument
that generalizes those results to the whole frozen phase and predicts
that the typical number of relevant elements scales as $\sqrt N$ on
the critical line. Though this argument is based on some
approximations which we can not control, its results
coincide for $K=1$ with the exact results by Flyvbjerg and Kjaer.

Let us suppose that we add a new element to a system with $N$ elements, $R$ of
which are relevant, while $S$ are stable and $I=N-R-S$ are indifferent,
{\it i.e.} neither stable nor relevant. The probability that the new element
is relevant can be computed as a function of $R$ and $S$, within some
approximations that we are going to discuss in a while. This probability is
equal to the fraction of relevant elements in the system with $N+1$
elements, given that the relevant elements are $R$ and the stable ones are
$S$ in the system with $N$ elements. We can then average over $R$ and
$S$ in order to get an equation connecting $r_{N+1}=\la R\ra_{N+1}/(N+1)$
to the moments of the distribution of $R$ in the system with $N$
elements. Since in the frozen phase and on the critical line $r_N$
vanishes, it will be enough to consider the first two moments of the
distribution, and the resulting equation can be solved asymptotically
in $N$.

The weakness of this approach lies on the assumptions that allow us to
express the probability that the new element is relevant as a function
of $R$ and $S$, as it will become soon clear.
We compute now this probability. To this aim, we need two steps:

\begin{enumerate}
\item As a first step, we have to extract the $K$ control elements and the
response function of the new element. As a consequence, the
new element can be stable, unstable or, if it receives an input from itself and
this input is relevant in the sense discussed above, relevant. 
The evaluation of the stability is perfectly equivalent to the mean field
argument, but this stability is only temporary because it can be altered by the
second step described below. Thus we call a new element that is stable
(unstable) after the first step a {\it temporarily} stable (unstable) element.

\item Then we have to send to the old system the signal of the new
  element. For each of the $KN$ old control connections we have a probability
$1/(N+1)$ that the connection is broken and the old control element is
substituted by the new element. This step perturbs the elements that
  control the new element and modifies its temporary stability. We
  have no chance to take this into account, unless we use some drastic
  approximations.
\end{enumerate}

In the second step, three situations can occur.

\begin{enumerate}
\item If the new element was relevant in the first step, the new step can not
modify this condition.
\item If the new element was unstable, it cannot become stable through the
feedback of its signal. So it will be relevant or indifferent, depending on
whether it sends an input to at least one relevant element or not.
\item If the new element was stable, its signal can destabilize some of the
elements that control it and thus it can become relevant through a feedback
mechanism, very hard to investigate analytically.
\end{enumerate}

To compute the probability of case 3, we should know the organization
of the network in the very detail and not only the number of relevant
and stable elements. We propose to bypass this difficulty considering
a different event: we will consider the new element relevant if it
receives a signal from a previously relevant element or from
itself. This is the simplest way to get a closed equation for the
average number of relevant elements. In this way we make two errors
of opposite sign: on one hand we overestimate the probability that a
temporarily unstable element becomes relevant, on the other one we
underestimate the probability that the new element is temporarily
unstable and we neglect the probability that a temporarily stable
element becomes relevant through a feedback loop. 

We think that this method captures at least the qualitative behavior of
the number of relevant elements. We have then to compare the estimate
given by this approximation to the simulations, because the
approximation is not under control. We present this argument because
its results agree with both the numerical results and with the
analytical calculations for $K=1$ and because we believe that it is
possible to improve this method and to keep the approximation under control.

Since we are interested in the frozen phase, where the fraction of unstable
elements vanishes in the infinite size limit, we can neglect the
eventuality that the new element is controlled by more than two
elements that were relevant in the old system. The results are
consistent with this assumption.
With these approximations we obtain the following equation for the
probability that the new element is relevant:

\bea \la r\ra_{N+1}=\sum_{n=0}^N \Pr\l\{R_N=n\r\}
\l[K\rh\l({n+1\over N+1}\r) \l(1-{n+1\over N+1}\r)^{K-1}\r.\label{rilev} \\
 +\l.\rh_2{K\choose 2}\l({n+1\over N+1}\r)^2
\l(1-{n+1\over N+1}\r)^{K-2}\r], \nonumber\eea
where $\rh_2$ represents the probability that a Boolean function of two
arguments is not constant and in terms of $\rh$ is given by
\be \rh_2=1-p^4-(1-p)^4=\rh\l(2-{\rh\over 2}\r). \ee

In the frozen phase it is sufficient to consider that the new element receives
only one signal from the previously relevant elements. So, posing $c=K\rh$,
the equation for the new fraction of relevant elements, $r$, is
\be \l\la r\r\ra_{N+1}\approx c\l\la r\r\ra_N+{c\over N}. \ee

The first term represents a new element that receives a relevant signal from
one of the previously relevant elements, the second term represents a new
element that receives its own relevant signal.

Thus the average number of relevant elements is independent on $N$ and its
asymptotic value is
\be \l\la R\r\ra_N={c\over 1-c}.\label{froz} \ee

This number diverges on the critical line $c=1$. In this case, we have to
consider also the eventuality that the new element receives a signal from two
of the previously relevant elements. Expanding to the second order in
$r=R/N$, and using the fact that $\rh_c=1/K$, we get the equation
\be \l\la r\r\ra_{N+1}\approx \l\la r\r\ra_N-\l({K-1\over 4K}\r)
\l\la r^2\r\ra_N+{1\over N}, \ee
whence, in the asymptotic regime where the variations of $\la r\ra_N$ are of
order $r/N$, we finally get
\be \l\la r^2\r\ra_N\approx\l({K-1\over 4K}\r){1\over N}. \ee

This means that the scale of the number of relevant elements grows, on the
chaotic phase, as $\sqrt N$.

We stress here that these computations are valid because of the finite
connectivity of the system. If we perform the limit $K\ri\infty$ on the above
result, we get that the scale of the number of relevant elements grow as
$1/2\sqrt N$. If, instead, we apply the limit $K\ri\infty$ prior to the limit
$N\ri\infty$ we get the trivial critical point $\rh=0$, where all the elements
are stable after one time step, while for every other $\rh$ value all the
elements are relevant.
Thus, the two limits do not commute. In fact, the equation (\ref{stable}) for
the fraction of stable variables and all the computations performed in this
section are valid only if we can neglect that the same element is chosen more
than once to control a given element, {\it i.e.} for $K\ll N$.

\vspace{0.5cm}
The result (\ref{froz}) coincides for $K=1$ with the analytical
computation by Flyvbjerg and Kjaer \cite{FK}, thus suggesting that the
distribution of relevant elements is independent on $N$ in the whole
frozen phase, and depends on the two parameters $K$ and $\rh$ only
through their product. This picture agrees with the results of the
annealed approximation, which predicts that the distribution of the
number of different elements in two asymptotic configurations is
independent on $N$ and depends only on the product of the parameters
$K$ and $\rh$ in the frozen phase \cite{BP0}.

Our simulations confirm that on the critical line the number of
relevant elements scales as $\sqrt N$ (see figure \ref{fig_ril4}). Also
the annealed approximation is consistent with this result, since it
predicts that the number of elements whose state is different in two
asymptotic configurations has to be rescaled with $\sqrt N$ on the
critical line \cite{BP0}. On the other hand the number of unstable
elements grows much faster with $N$  (numerically it is found that it
goes as $N^{3/4}$, see below) but this discrepancy is only apparent,
since the asymptotic Hamming distance is related more to the number of
relevant elements than to this quantity.

For later convenience (see our next paper) it is also interesting to
compute the effective connectivity, defined as the average value of the
relevant connections between relevant elements. Let us compute it by
imposing the condition that the network has $R$ relevant elements.

The effective connectivity is equal to the average number of connections
between the new element and the other relevant elements of the older system,
with the condition that the new element is relevant. From equation
(\ref{rilev}) we have, at the leading order in $R/N$:

\bea \label{Ceff} K_{eff}(R)=&&
{c\RN \l(1-\RN\r)^{K-1}+2\rh_2{K\choose 2}\l(\RN\r)^2\l(1-\RN\r)^{K-2}\over
c\RN \l(1-\RN\r)^{K-1}+\rh_2{K\choose 2}\l(\RN\r)^2\l(1-\RN\r)^{K-2}}\\\
\nonumber &&\approx 1+A(K,\rh)\RN. \eea

This equation shows that the effective connectivity minus 1 goes to
zero as $R/N$ in the frozen phase (where $R/N\propto 1/N$) and on the
critical line (where $R/N\propto 1/\sqrt N$). For a fixed system size,
the effective connectivity increases linearly with the number of
relevant elements. 

\section{Numerical results}
\subsection{Magnetization and stable elements}
As a first step, our algorithm has to identify the stable elements. It
does this by measuring their magnetization. We thus discuss our numerical
results starting from this quantity.

The magnetization $m_i^\a$ of element $i$ on the cycle $\Gamma_\a$ can
be defined as the average activity of the element along the cycle:
\be m_i^\a={1\over L_\a}\sum_{C\in \Gamma_\a}\s_i(C). \label{mag}\ee

\begin{figure}
\centering
\epsfysize=12.0cm 
\epsfxsize=15.0cm 
\epsffile{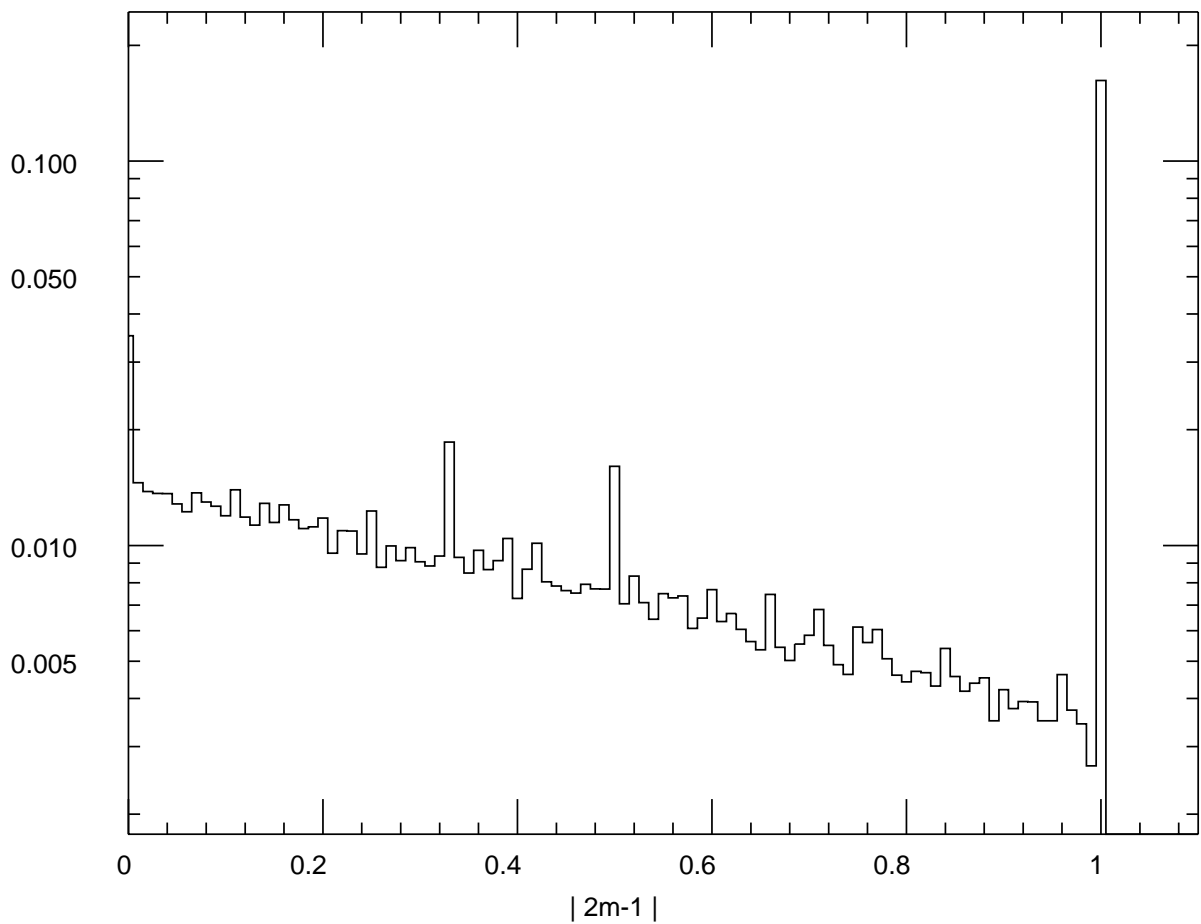}
\caption{\it The histogram of $\mid 2m-1\mid$, where $m$ is the
  average activity of a given node in a single cycle of a given
  network, is shown for $K=3$ and $N=75$. The peaks correspond to
  simple rational values of the argument: for instance, $1/8, 1/7,
  1/6$... $800$ sample networks were generated, and $500$ trajectories
  simulated on each of them.}
\label{fig_magneti}
\end{figure}

The distribution of this variable, shown in figure \ref{fig_magneti}
for $K=3$ and $N=75$, has many peaks, corresponding to simple
rational values. This perhaps reflects the fact that the
relevant elements are divided into asymptotically independent modules,
so that a cycle can be decomposed into several independent shorter
cycles. This subject will be further discussed in our second paper.

Our results have to be compared to the analytical work by Derrida and
Flyvbjerg \cite{DF3}. 
They defined the magnetization of element $i$ at time $t$ on a given network
as the activity of the element at time $t$ averaged over many initial
configurations and could compute analytically its stationary
distribution, in the limit $N\ri\infty$, using
the annealed approximation, that can be shown to be exact for this purpose.
The picture they got is different from ours, in particular we see
peaks much higher than theirs. For instance the peak at $\mid 2m-1
\mid =1$, which gives information on the size of the stable core of
the network, is about 10 times larger then expected, and the first
moments of the magnetization, that can be computed analytically, are
larger than the predicted values. Thus we performed other simulations
that strongly suggest that these discrepancies are finite size
effects, and we present an argument that explains their origin.

\begin{figure}
\centering
\epsfysize=12.0cm 
\epsfxsize=15.0cm 
\epsffile{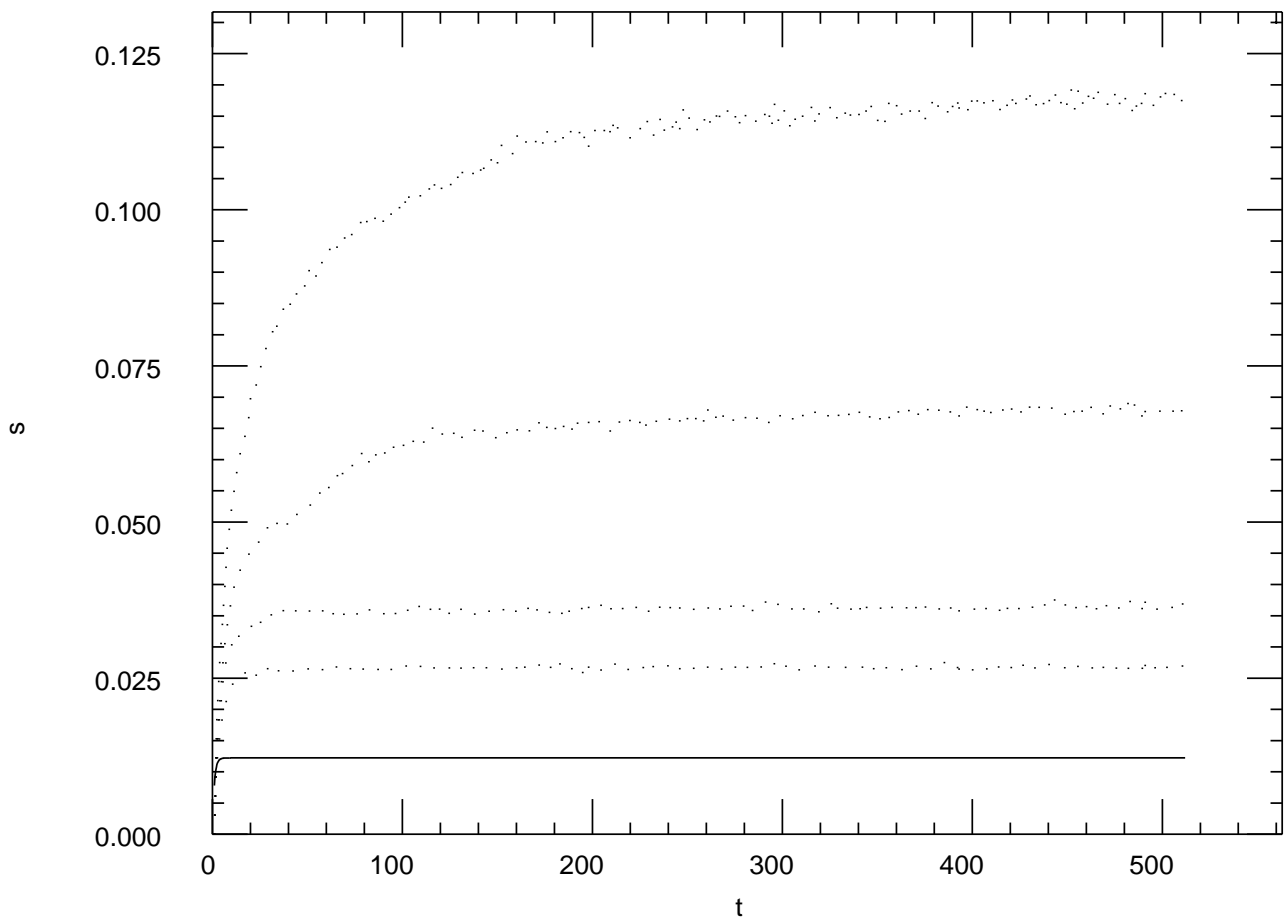}
\caption{\it Fraction $s$ of nodes which are at time $t$ in the same state in
all the 200 different trajectories simulated. Data are average over 200 sample
networks. $K=3$, $N=50$, 100, 150 and 200 ($N$ grows from top to bottom). The
solid line is the mean field equation, 
$s(t+1)=\sum_{l=0}^K {K \choose l}s(t)^{K-l}\left(1-s(t)\right)^l
{1\over 2^{2^l-1}}$.}
\label{fig_score}
\end{figure}

In order to investigate larger systems we had to change the definition
of the magnetization. The definition (\ref{mag}) is numerically
cumbersome, since the measure takes place only after that a cycle has
been found, and this means, for chaotic systems, that we have to wait
a time exponentially increasing with $N$. Thus we neglected this
condition and we measured the magnetization of the variable $i$ at
time $t$ as the average activity  of the variable with respect to
different initial conditions (this definition coincides with the one
used by Derrida and Flyvbjerg). For very large $t$, when all
trajectories have reached a limit cycle, this quantity tends to the
asymptotic value 

\be m_i=\sum_\a W_\a m_i^\a,\label{magna}\ee
where $W_\a$ is the weight of the basin of cycle $\Gamma_\a$ and
$m_i^\a$ is defined in (\ref{mag}). We observed that $m(t)$
reaches a stationary value (within some precision) much earlier than the
typical time at which the trajectories reach their limit cycles. At
first sight surprisingly, the time after which $m(t)$ reaches its stationary
distribution does decrease with system size instead of increasing
(see figure \ref{fig_score}).

We measured the second and fourth moment of the magnetization in a
system with $K=3$ and $\rh=1/2$, and we
found a large positive correction to the infinite size values computed
by Derrida and Flyvbjerg \cite{DF3}. The values found coincide
within the statistical error with those obtained from equation
(\ref{mag}) for a system of small size for which we did an explicit
comparison. These values can be fitted to the sum of the infinite size
value, that we got from \cite{DF3}, plus an exponentially decreasing
term. The exponent of the best fit turned out to be the same for both
the moments that we measured: we found
$m_2(N)\approx 0.236+0.24\cdot \exp(-N/70)$, and
$m_4(N)\approx 0.128+0.26\cdot \exp(-N/70)$.

\vspace{0.5cm}
The measure of the magnetization allows to identify the stable
elements as the elements with $\sum_\a W_\a m_i^\a$ equal either to 0
or to 1. The two definitions of the magnetization gave roughly the same
number of stable elements in the cases where we could compare the
results, but with the second method we could consider much larger
systems (we recall that the difference between the
two methods is that in the first case a cycle  has been reached while
in the second one the system is still in some transient
configuration). The second method was used only to study finite size
effects, since it does not allow to identify the
relevant elements (see below).

Both the methods overestimate the number of stable elements, since
it could happen that an element appearing stable in our sample of
trajectories (some hundreds) oscillates in a cycle that is not reached
by any of them. We checked that the results do not change qualitatively
if we consider a larger number of trajectories. 

The fraction of stable nodes measured in simulations with $K=3$
and $N$ ranging from 50 to 200 have been compared to the prediction of the
mean field theory by Flyvbjerg. The networks with $N=50$ have a stable core
about 10 times larger than the mean field value (in this case we
measured the magnetization using both the above definitions, while for
larger systems only equation (\ref{magna}) was used). The corrections
to the mean field value, that is exact in the infinite size limit,
appear to decay exponentially with a rate identical, within
statistical errors, to the decay rate of the corrections to the moments of the
magnetization: we found
$s(N)\approx 0.0122+0.21\cdot \exp(-N/70)$. 

For every size of the systems which we simulated the stable core is
then much larger than it would be in an infinite system.
On this ground, we may expect very important finite size effects concerning
the dynamical properties of the system.

\vspace{0.5cm}
Summarizing, the distribution of the magnetization for finite systems
has the following characteristic: 1) The asymptotic value is reached
after a time that {\it decreases} with system size; 2) the
corrections to the infinite size values are very large; and 3) these
corrections decrease exponentially with system size.
These apparently strange finite size effects have a simple
interpretation: they arise as a consequence of the periodic dynamic of
the random networks. 

The mean field values of the magnetization and of the
stable core are computed within the annealed approximation without
taking into account the fact that the asymptotic dynamic is
periodic. As we proposed in \cite{BP0}, the existence of limit cycles must
be taken into account in the framework of the annealed approximation
in this way: if at time $t$ all the configurations generated are
different ({\it i.e.} the trajectory is still open) we treat the
quantities of interest (distance, magnetization or stable core) as a
Markovian stochastic process; if one configuration has been found twice
(the trajectory is closed) we impose the condition that all
quantities are periodic. Thus the master equation for the distribution
for the number of stable variables is, in the framework of the annealed
approximation:

\bea
& & \Pr\l\{ S(t+1)=S',O(t+1)\mid S(t)=S, O(t)\r\}=
{N\choose
  S'}\l(\g(s)\r)^{S'}\l(1-\g(s)\r)^{N-S'}\l(1-\pi_N(S,t)\r)
\nonumber\\
& & \Pr\l\{ S(t+1)=S',\overline{O(t+1)}\mid S(t)=S, O(t)\r\}=
{N\choose S'}\l(\g(s)\r)^{S'}\l(1-\g(s)\r)^{N-S'}\pi_N(S,t)
\nonumber\\
& & \Pr\l\{ S(t+1)=S',\overline{O(t+1)}\mid S(t)=S,
\overline{O(t)}\r\}= \delta_{SS'}, \label{stable2}
\eea
where $S(t)$ is the number of stable elements, $s=S/N$, $O(t)$ stands for
the condition that the trajectory is open at time $t$ (no
configuration has been visited twice), $\overline{O(t)}$ stands for
the condition complementary to $O(t)$ (the trajectory has closed on a
previously visited configuration) and
$\pi_N(S,t)$ is the probability that a trajectory open
at time $t$ and with $S$ stable elements at time $t+1$ closes at that
time. Finally, $\gamma(s)$ is given by equation (\ref{stable}). 

We don't know how to compute $\pi_N(S,t)$, but it is clear that this
is an increasing function of $S$ for fixed $t$: the more elements are
stable, the more it is likely that the trajectory closes. The
infinite size value of the stable core is given by equation
(\ref{stable}), that represents the evolution of the most probable
number of stable variables. It is clear that the corrections to this value
are positive, and that they go to zero as soon
as the closing time becomes much larger than the time necessary for
the stable core to reach its stationary value in an infinite system
(where all trajectories are still open). Thus we expect that these
corrections vanish as a power law of the typical length of the cycles:
in the chaotic phase this means that the finite size corrections due to this
effect vanish exponentially with system size, as we observed
simulating systems with $K=3$ and $\rh=1/2$. Lastly, this argument
implies that the time after which the distribution of the stable
elements becomes stationary is shorter in an infinite system than in a
small system, where the evolution of $S(t)$ is coupled to the closure
of the periodic orbits. Thus the correction of the annealed
approximation to take into account the existence of periodic
attractors can account for all the features of the finite size effects
that we observed.

\subsection{The relevant elements in the chaotic phase}

After having identified the stable elements we detect the relevant
elements using the algorithm described in the second section and we
study how this quantity influences the dynamical properties
of the network. The main results are that the average
cycle length grows almost exponentially with the number of relevant
variables in some range of this variable and the average weight of the
attraction basins has apparently a non monotonic behavior versus the number of
relevant variables. This qualitative features are the same both in the
chaotic phase and on the critical line, but the ranges of $R$ in which
these things happen are quite different in the two cases. We start
discussing the situation in the chaotic phase.

The simulations were done generating at random $20,000$ sample
networks and running 200 trajectories on each of them. The parameters
considered in this section are  $K=3$, $\rh=1/2$ and
system size $N$ ranging from 30 to 60 elements.

\begin{figure}
\centering
\epsfysize=12.0cm 
\epsfxsize=15.0cm 
\epsffile{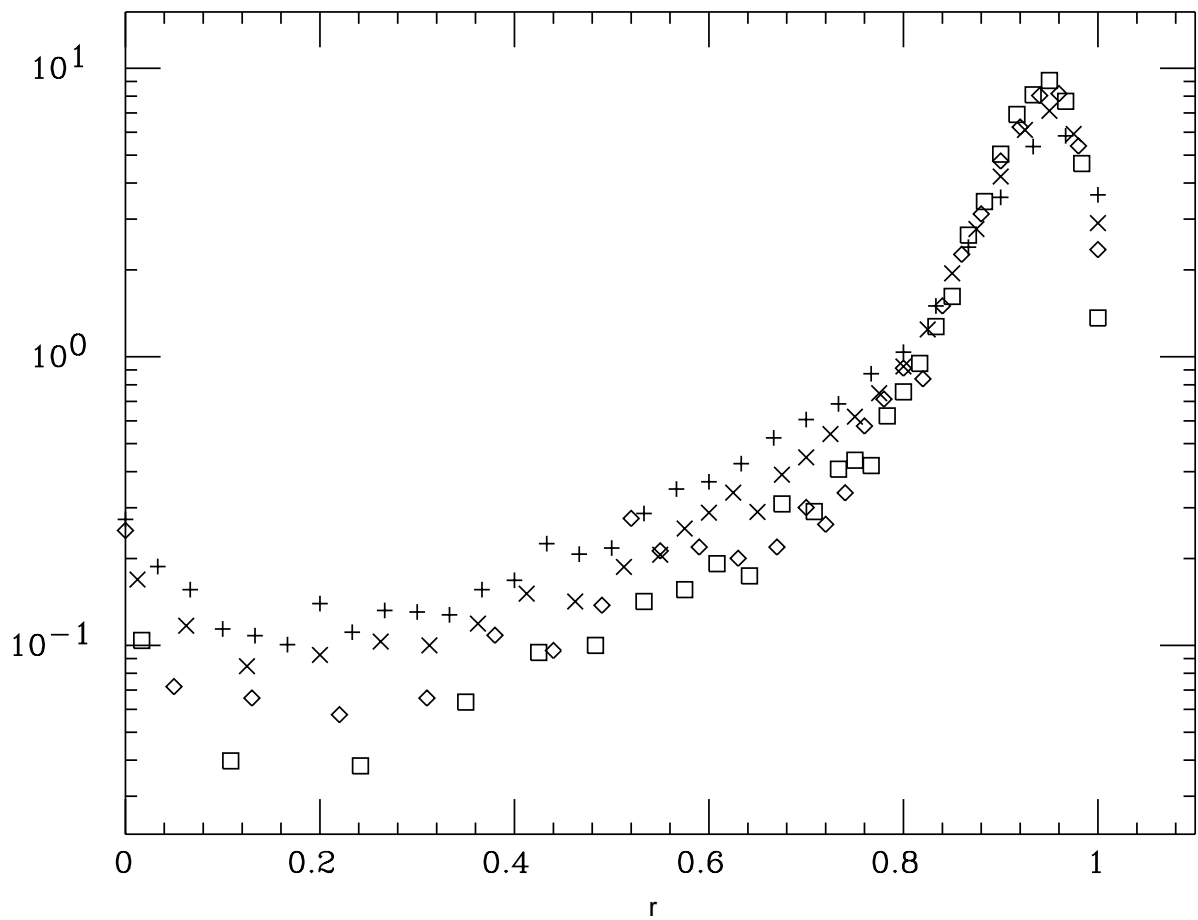}
\caption{\it Histogram of the networks with a fraction $r$ of relevant
  elements. $K=3$, $N=30$ (+), 40 (X), 50 ($\diamondsuit$) and 60
  (*). The networks generated were $20000$, the relevant elements were
  identified simulating 200 trajectories on each of them.}
\label{fig_ril}
\end{figure}

Figure \ref{fig_ril} shows the density of the distribution of the fraction
$r_N$ of relevant variables, $r_N=R/N$. The density relative to the
most probable value increases with system size, and it appears that
$r_N$ tends to be delta-distributed in the infinite size limit, as it is
expected on the ground of the annealed master equation (\ref{stable2}).
We observe an excess of networks with very few relevant
elements ({\it i.e.} very many stable elements), consistently with the
finite size effects discussed in last section. This excess seems to
disappear in the infinite size limit.

\begin{figure}
\centering
\epsfysize=12.0cm 
\epsfxsize=15.0cm 
\epsffile{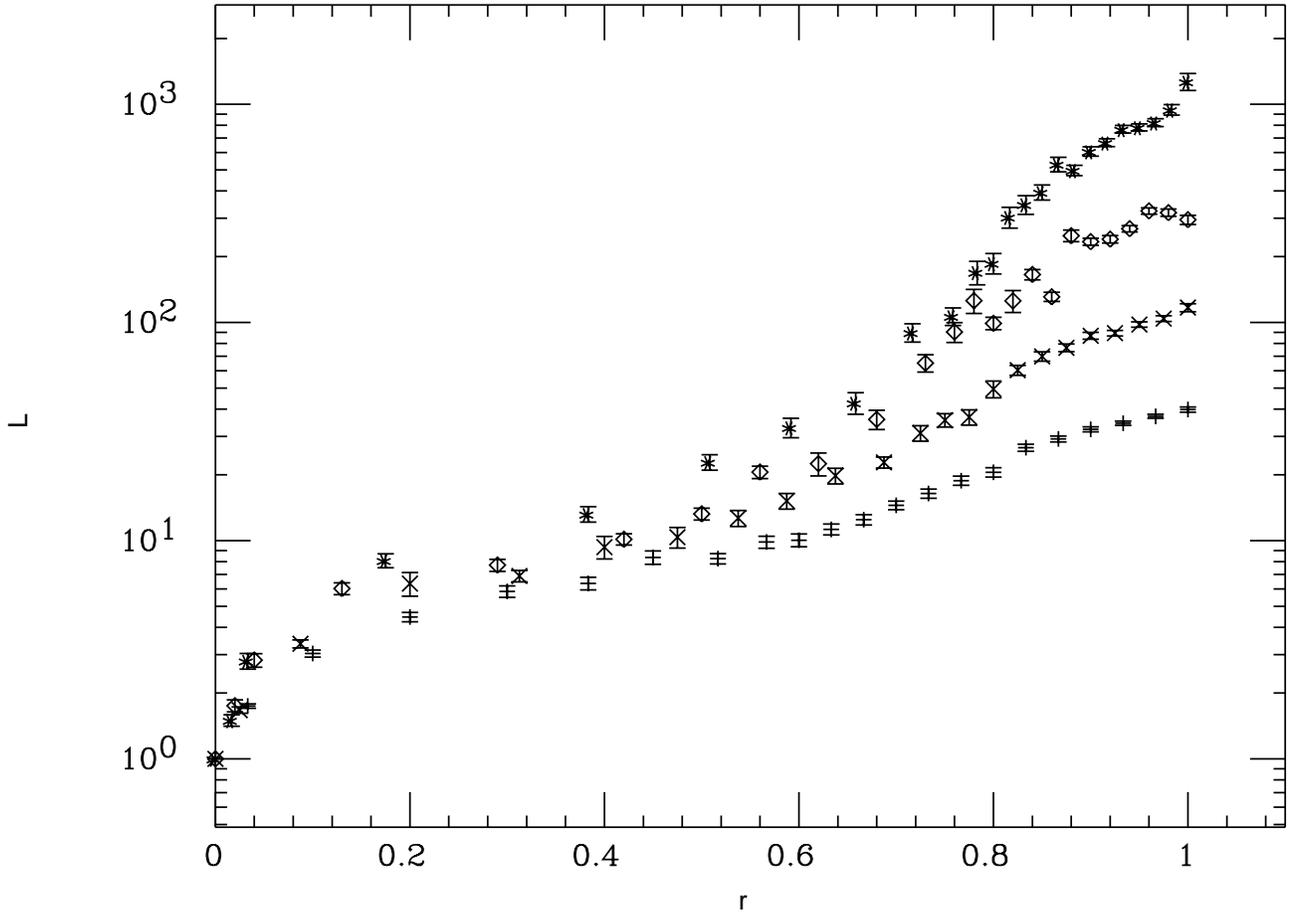}
\caption{\it The average period in networks with a fraction $r$ of relevant
nodes. $K=3$, $N=30,40,50$ and $60$. Each curve was obtained generating at
random $20000$ networks and simulating $200$ trajectories on each of them.}
\label{perril}
\end{figure}

Then we show the average length of the cycles in networks with $R$
relevant elements (figure \ref{perril}). 
This quantity increases almost exponentially with $R$ when
$r=R/N$ is large, while its behavior is different for small $r$. The
crossover takes place at about $r=0.5$. Thus the number of relevant
elements turns out to have a very important influence on the typical
length of the cycles

We have also measured the conditional distribution of the length of
the cycles in networks with $R$ relevant elements. When $R$ is close
to $N$ the distribution decays as a stretched exponential with an
exponent smaller than one, very close to the one found in the
unconditioned distribution. Thus the deviation of the unconditioned
distribution from the prediction of the annealed approximation, that
predicts a much narrower distribution, is not a consequence of the
existence of the relevant elements.

The other quantity that we measured is the average weight of
the attraction basins,  $Y_2$, defined by the equation

\be Y_2=\sum_\a W_\a^2, \ee
where $W_\a$ is the attraction basin of cycle $\a$. We used the method
proposed by Derrida and Flyvbjerg \cite{DF1}, that is based on the
fact that $Y_2$ is
equal to the probability that two trajectories chosen at random
end up on the same attractor.

From our data (not shown) it appears that $Y_2$ has a non monotonic behavior
as a function of $r$: for very small $r$ it decreases from the value 1,
corresponding to $r=0$, reaches a minimum and rapidly increases. At
large $r$, $Y_2$ does not seem to be correlated with $r$ (at least
within the statistical error, that is rather large). We will
see in the next paper that the decreasing behavior at small $r$ can be
interpreted as an effect of the modular organization of Kauffman networks.

\subsection{Relevant elements on the critical line}

We simulated systems with $K=4$ and the critical value
$\rh=1/4$. Systems size ranges from 120 to 1600. Concerning the
statistical properties of the attractors, these networks have a
behavior very similar to that of the more studied $K=2$, $\rh=1/2$
networks \cite{BP1}.

In these networks, the number of relevant elements appears to scale as
$\sqrt N$, in agreement with the argument presented in section 4. The
number of unstable elements, on the other hand, appears
to scale as $N^{3/4}$. This implies that the probability to extract
at random an element which is relevant, scaling as $N^{-1/2}$, is
approximately proportional to the square of the probability to extract
at random an element which is relevant ($N^{-1/4}$). 

\begin{figure}
\centerline{\psfig{figure=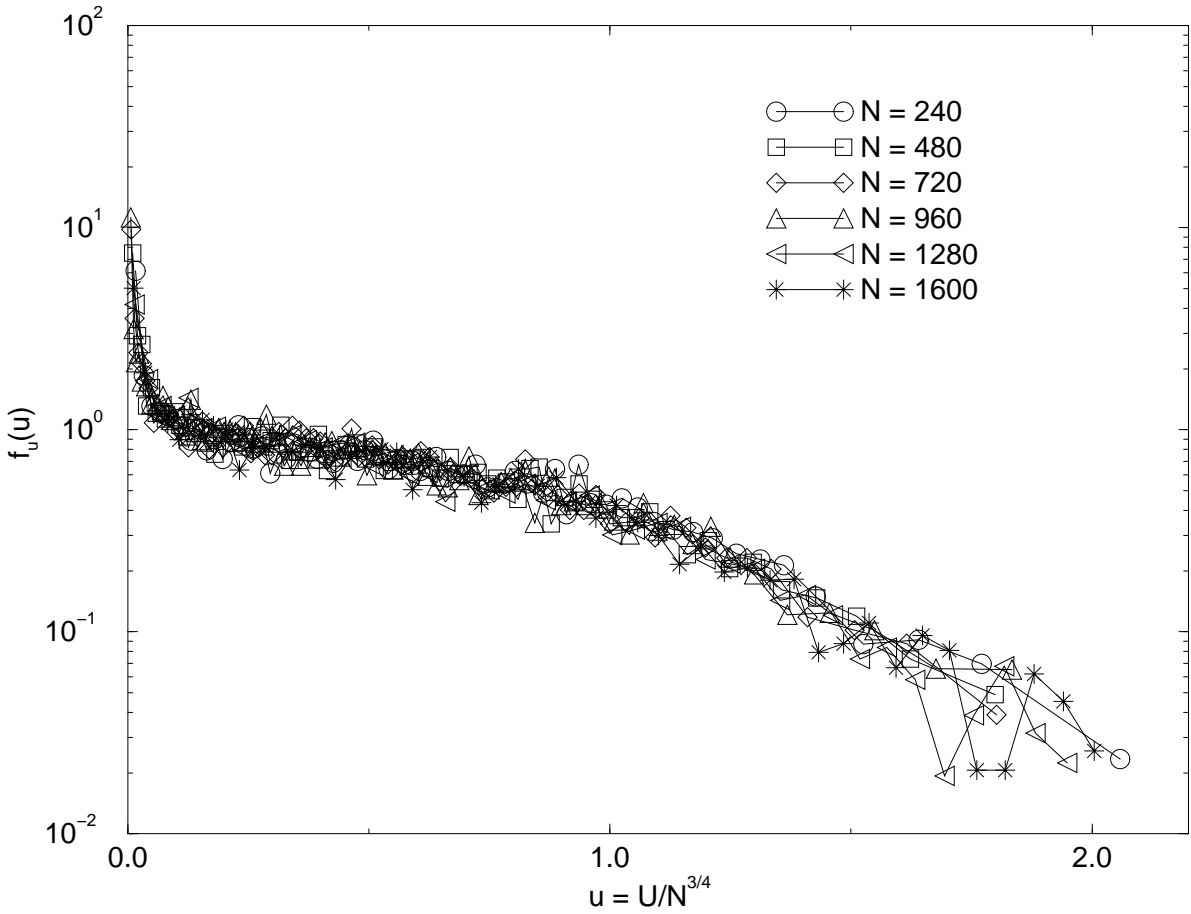,height=12cm,angle=-90}}
\caption{\it  Probability density of the rescaled number of unstable
  elements $x_u=U/N^{3/4}$, where $U$ is the number of unstable
  elements in a random network. The systems are at the critical point
  $K=4$, $\rh=1/4$ and system size ranges from 240 to 1600.}
\label{fig_unst}
\end{figure}

These scaling laws can be observed both looking at the average quantities
and looking at the whole distribution. The average number of unstable
variables is found to follow the power law $U\propto N^a$, with
$a=0.74\pm 0.01$. We then define the rescaled
variable $x_u=U/N^{3/4}$, and we compare its probability density
for various system sizes. As it can be seen in figure
\ref{fig_unst} the different curves superimpose within the
statistical errors. This suggests that $x_u$ has a well defined
probability density in the infinite size limit, although our data are
rather noisy to state this point without doubts. We can distinguish in
the distribution three different ranges with different
characteristics: at vanishingly small values of $x_u$ (ranging from
$U=0$ up to $U=4$) the density decreases very fast.
At intermediate values, roughly up to $x_u=1$, it looks to decrease
approximately as a power law with a small exponent (the best fit exponents
that we found range from 0.25 to 0.40, showing some tendency to
increase with system size). Asymptotically, for large $x_u$, the best
fit is a stretched exponential, $f(x)\approx \exp\l(-Cx^\b\r)$, with an
exponent compatible with $\b=1.7\pm 1$ for all the systems that with
studied with $N$ larger than 240.

\begin{figure}
\centerline{\psfig{figure=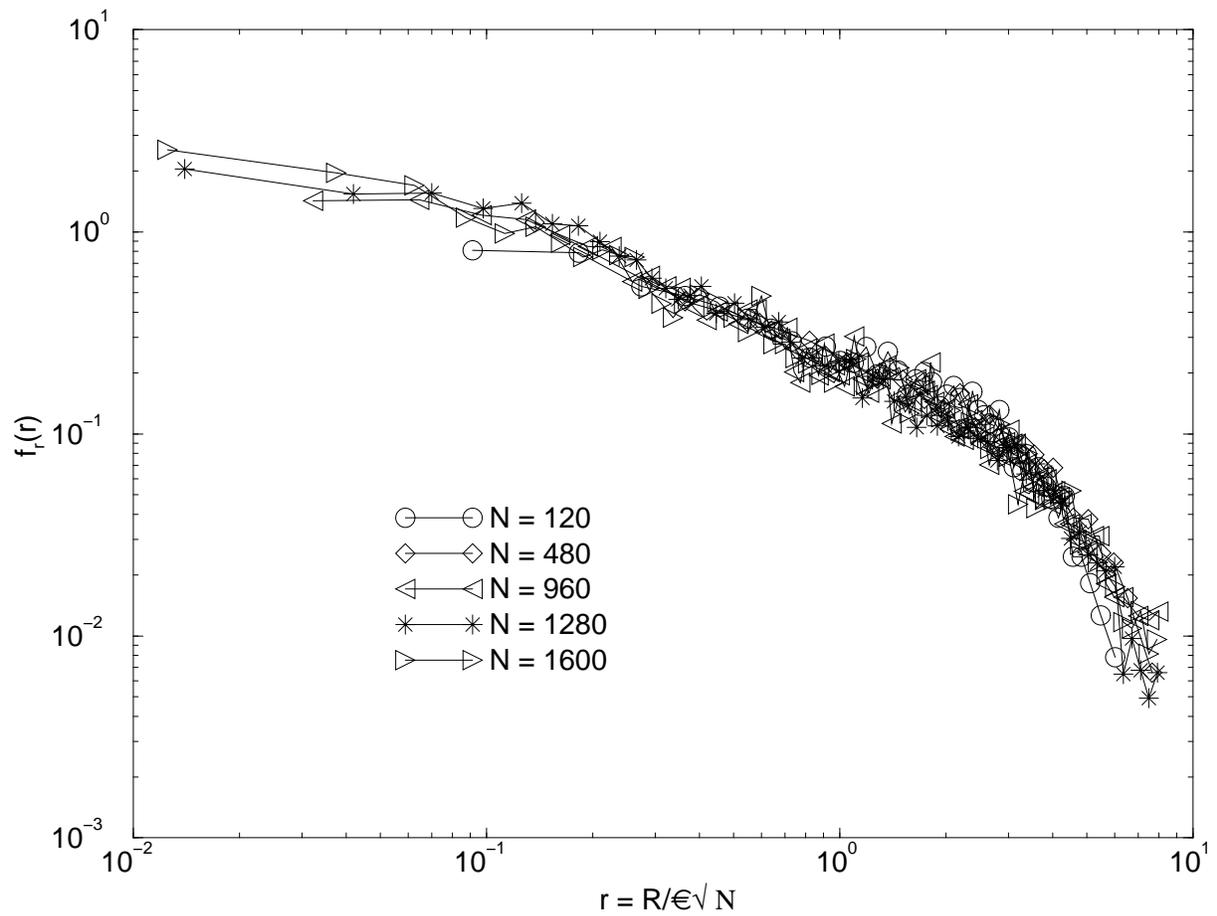,height=12cm,angle=-90}}

\caption{\it  Probability density of the rescaled number of relevant
elements $x_r=R/N^{1/2}$, where $R$ is the number of relevant elements in a
random network. The systems are at the critical point $K=4$, $\rh=1/4$ and
system size ranges from 120 to 1600.}
\label{fig_ril4}
\end{figure}

The number of relevant variables was studied in a similar way. Its
average value increases as a power law of $N$, $\la R\ra\propto R^b$, with
$b=0.52\pm 0.02$.  The rescaled variable $x_r=R/\sqrt N$ looks to
have a well defined distribution in the infinite size limit, as it is
shown in figure \ref{fig_ril4}, where the probability density of $x_r$
is plotted for system sizes ranging from 120 to 1600. 
For large $x$ the density of the distribution is well fitted by a
stretched exponential, $\exp\l(-Cx^\b\r)$, with the exponent $\b$
compatible with the value $\b=0.56\pm 0.02$ for system size larger
than 240.

\begin{figure}
\centerline{\psfig{figure=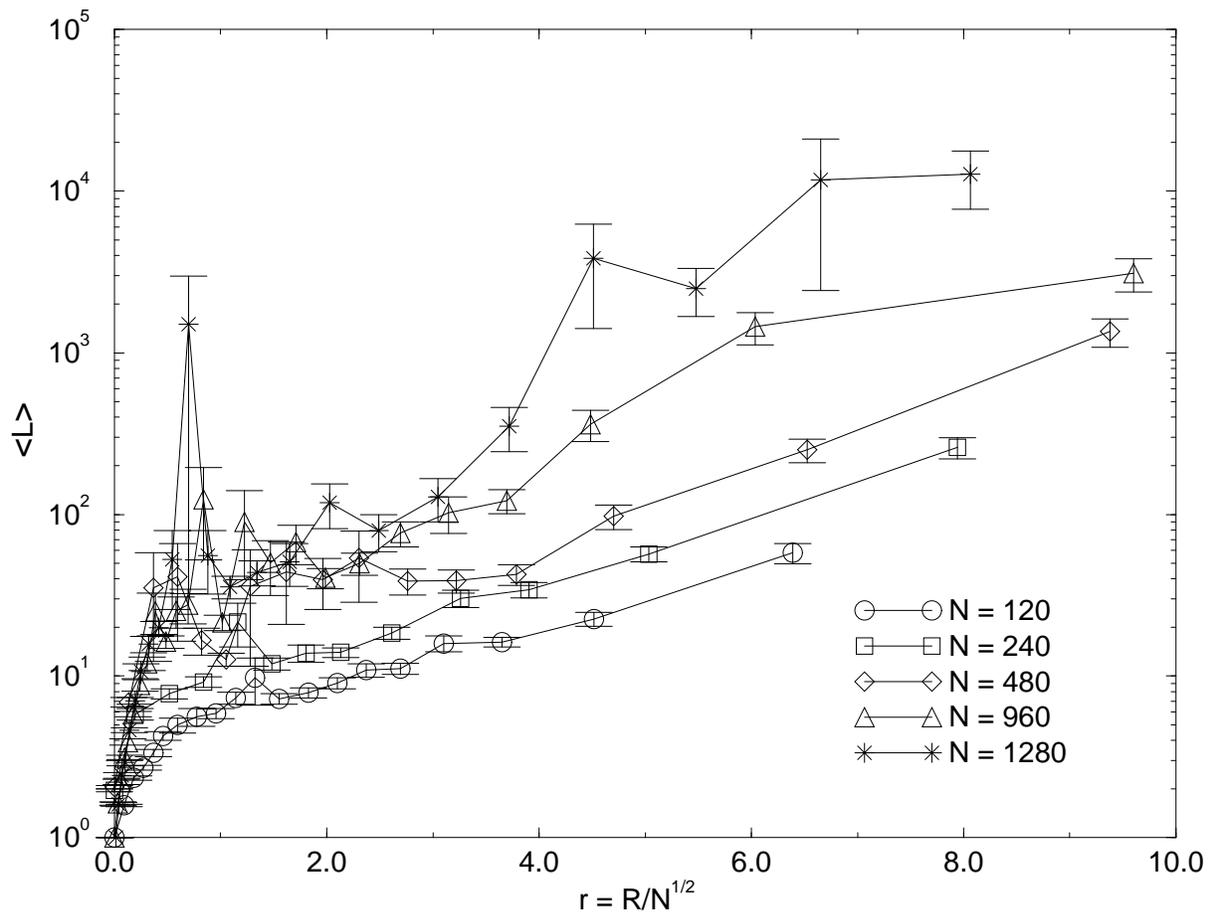,height=12cm,angle=-90}}

\caption{\it  Average cycle length in networks with $R$ relevant elements as a
function of the rescaled variable $r=R/N^{1/2}$, for critical systems
($K=4$, $\rh=1/4$) of size ranging from 120 to 1600.}
\label{fig_perril4}
\end{figure}

\vspace{0.5cm}
The average length of the cycles increases exponentially as a function of the
number of relevant elements, for $r$ large, and more slowly for $r$
small, just as it happens in the chaotic phase. Figure
\ref{fig_perril4} shows on a logarithmic scale the behavior of the
average length of the cycles as a function of the rescaled number of relevant
elements, $x_r=R/\sqrt N$, for different system sizes at the critical
point $K=4$, $\rh=1/4$.

\begin{figure}
\centerline{\psfig{figure=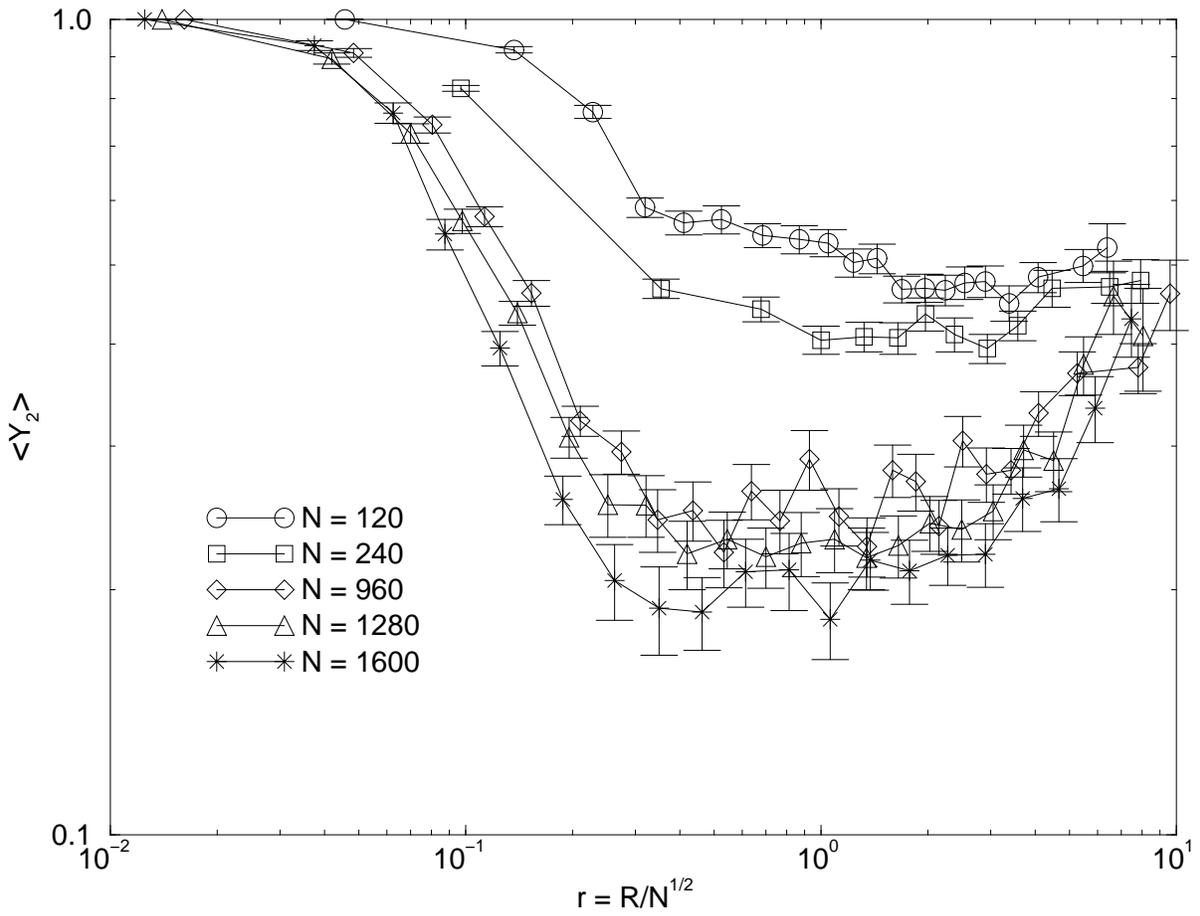,height=12cm,angle=-90}}

\caption{\it  Average value of $Y_2$ in networks with $R$ relevant elements,
as a function of the rescaled variable $r=R\/sqrt N$, for critical systems
($K=4$, $\rh=1/4$) of size from 120 to 1600.}
\label{fig_yril4}
\end{figure}

The average weight of the attraction basins, $Y_2$, has a non
monotonic behavior as a function of the number of relevant elements,
as it happens in the chaotic phase. The value of $Y_2(R)$ is one for
$R=0$, then decreases to a minimum value and increases very slowly, as
it is shown in figure \ref{fig_yril4}, where $\overline{Y_2}$ is
plotted against $x_r=R/\sqrt N$, for $K=4$, $\rh=1/4$ and different
system sizes. 

Nevertheless, there are two important
differences with respect to the chaotic phase: first, the range where
$Y_2(R)$ is a decreasing function is much wider on the critical line than in
the chaotic phase; then, on the critical line the curves corresponding to a
smaller $N$ value are lower, while in the chaotic phase the contrary
holds. As a consequence, if we average $Y_2(R)$ over $R$ on the
critical line, we get a quantity vanishing in the infinite size
limit \cite{BP1}, while the average weight of the attraction basins is
finite and very close to the Random Map value in the chaotic phase \cite{BP0}.

This difference and the non monotonic $R$ behavior of $Y_2(R)$ have
a clear interpretation in the framework of the modular
organization of Kauffman networks \cite{BP3}. We thus postpone to that
paper the discussion of our results.

%
%


\vfill\eject


\end{document}